\documentclass[aps,prl,showpacs,reprint,gropedaddress]{revtex4-2}  
\usepackage{amsmath}
\usepackage{amsfonts}
\usepackage{graphicx}
\usepackage{braket}
\usepackage{float}
\usepackage{color}
\usepackage{bm}
\usepackage{amssymb}
\usepackage[english]{babel}

\begin{document}

\title{Geometric phase and wave--particle duality of the photon}

\author{Elvis Pillinen}
\email[]{elvis.pillinen@uef.fi}
\author{Atri Halder}
\author{Ari T. Friberg}
\author{Tero Set{\"a}l{\"a}}
\author{Andreas Norrman}

\affiliation{Center for Photonics Sciences, University of Eastern Finland, P.O. Box 111, FI-80101 Joensuu, Finland}

\date{\today}

\begin{abstract}
The concepts of geometric phase and wave--particle duality are interlinked to several fundamental phenomena in quantum physics, but their mutual relationship still forms an uncharted open problem. Here we address this question by studying the geometric phase of a photon in double-slit interference. We especially discover a general complementarity relation for the photon that connects the geometric phase it exhibits in the observation plane and the which-path information it encases at the two slits. The relation can be seen as quantifying wave--particle duality of the photon via the geometric phase, thus corroborating a foundational link between two ubiquitous notions in quantum physics research.

\end{abstract}

\maketitle

\emph{Introduction}.---The geometric phase~\cite{Berry1984}, which is the phase that a physical system acquires as it evolves along a curved trajectory in the underlying parameter space, is a universal concept within the physical sciences~\cite{Shapere1989}. It is encountered in particle physics, condensed-matter physics, fluid dynamics, and astrophysics, among other branches~\cite{Cohen2019}, and offers unique opportunities for emerging quantum technologies~\cite{Sjoqvist2015,Zhang2023}. Wave--particle duality is another central notion in modern physics that restricts the coexistence of interferometric “which-path information” (particle behavior) and fringe visibility (wave behavior) of quantum objects~\cite{Englert1996}. It is perhaps the most recognized manifestation of quantum complementarity~\cite{Scully1991,Zeilinger1999,Shadbolt2014} and has been observed in a wide variety of quantum physical systems, such as elementary particles~\cite{Zeilinger1988,Zou1991,Bucks1998}, atoms~\cite{Carnal1991,Durr1998a,Durr1998b}, molecules~\cite{Arndt1999,Juffmann2012}, and even antimatter~\cite{Sala2019}.

In optical physics, the geometric phase arises from the change in the light's polarization state~\cite{Pancharatnam1956,Berry1987,Hariharan2005,Bliokh2019,Cisowski2022} and it has found numerous applications in advanced light manipulation~\cite{Jisha2021}. Even a single photon can carry the geometric phase~\cite{Kwiat1991}, which can be seen as a specific wave facet of the photon. For classical light fields, the geometric phase was very recently observed in the continuous, periodic polarization-state pattern in two-slit interference~\cite{Hannonen2020,Leinonen2023}. The recognition that light has the ability to exhibit such interferometric polarization-state modulation has further revealed fundamental aspects of wave–particle duality of the photon~\cite{Norrman2017,Norrman2020}. These facts hint that the geometric phase is profoundly linked to the dual wave–particle nature of light at the single-photon level, but exactly how has remained an unresolved physical problem. 

In this Letter, we investigate the geometric phase of the photon in double-slit interference and show that it is directly connected, in a deeply complementary manner, to the photon's particle characteristics at the two slits. In particular, we formulate a fundamental wave--particle duality relation for the photon in terms of the geometric phase that it displays in the interference plane and the which-path information that it carries in the slit plane. Our work thus establishes a link between two elementary notions in physics and provides foundational  insights into the nature of the photon.

\begin{figure}[b]
\includegraphics[clip, width=.4 \textwidth]{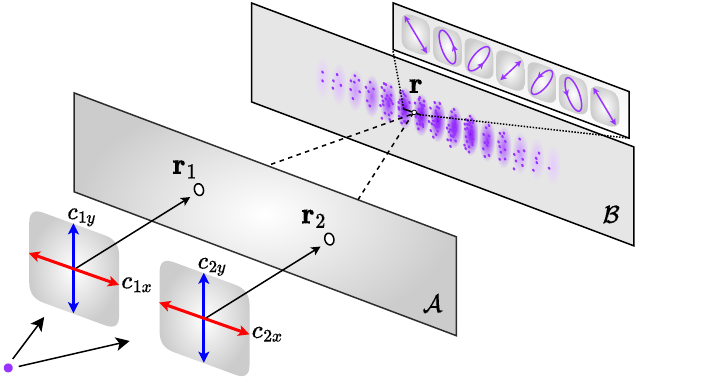}
\caption{System under study. A photon in a four-mode state $\ket{\Psi}=c_{1x}\!\ket{1,0,0,0}+c_{1y}\!\ket{0,1,0,0}+c_{2x}\!\ket{0,0,1,0}+c_{2y}\!\ket{0,0,0,1}$ strikes screen $\mathcal{A}$ with two small slits at positions $\mathbf{r}_1$ and $\mathbf{r}_2$. The interfering light is observed at position $\mathbf{r}$ in plane $\mathcal{B}$ where periodic intensity and polarization-state fringes appear after repeated experimental runs.}
\label{doubleslit}
\end{figure}

\emph{System under study}.---Let us consider monochromatic quantum light impinging on two identical slits (pinholes) located at $\mathbf{r}_1$ and $\mathbf{r}_2$ in an opaque screen $\mathcal{A}$ (see Fig.~1). The light emerging from the slits is observed on another screen $\mathcal{B}$ far from $\mathcal{A}$ at position $\mathbf{r}$ in the paraxial domain. Under these circumstances, the electric field operator at $\mathcal{B}$ is~\cite{Norrman2017,Mandel1995}
\begin{equation}
\label{E}
\hat{\mathbf{E}}(\mathbf{r})=K\Big[\hat{\mathbf{E}}(\mathbf{r}_{1})\frac{e^{ikr_{1}}}{r_{1}}+\hat{\mathbf{E}}(\mathbf{r}_{2})\frac{e^{ikr_{2}}}{r_{2}}\Big],
\end{equation} 
where $K$ is a constant, $k$ is the wave number in free space, and $r_{m}=|\mathbf{r}-\mathbf{r}_{m}|$ with $m\in\{1,2\}$. Moreover, each of the electric field operators $\hat{\mathbf{E}}(\mathbf{r}_{1})$ and $\hat{\mathbf{E}}(\mathbf{r}_{2})$ in the slit plane $\mathcal{A}$ contains two orthogonal polarization modes ($x$ and $y$), characterized by the annihilation operators $\hat{a}_{1x},\hat{a}_{1y}$ and $\hat{a}_{2x},\hat{a}_{2y}$. We are especially interested in the case where the light field is in an arbitrary, pure single-photon state
\begin{equation}
\label{photon}
\begin{aligned}
\ket{\Psi} =& \: c_{1x} \ket{1,0,0,0} + c_{1y} \ket{0,1,0,0} \\&+ c_{2x} \ket{0,0,1,0} + c_{2y} \ket{0,0,0,1}.
\end{aligned} 
\end{equation}
Here $\ket{n_{1x},n_{1y},n_{2x},n_{2y}}$ is a four-mode Fock state, with $n_{m\mu}$ denoting the photon number in the mode $\mu\in\{x, y\}$ at slit $m\in\{1, 2\}$, and $|c_{1x}|^{2}+|c_{1y}|^{2}+|c_{2x}|^{2}+|c_{2y}|^{2}=1$, with $|c_{m \mu}|^2$ giving the probability to find the photon in the corresponding mode.

The average intensity and polarization-state distributions of the one-photon light at $\mathcal{B}$ are completely specified by the four Stokes parameters~\cite{Brosseau1998}
\begin{equation}
\label{Stokes}
S_{j}(\mathbf{r})=\bra{\Psi}\!\hat{\mathbf{E}}^{\dagger}(\mathbf{r})\boldsymbol{\sigma}_{j}\hat{\mathbf{E}}(\mathbf{r})\!\ket{\Psi}, \ \ j\in\{0,1,2,3\},
\end{equation} 
where the dagger stands for the adjoint, $\boldsymbol{\sigma}_{0}$ is the $2\times2$ identity matrix, and $\{\boldsymbol{\sigma}_{1},\boldsymbol{\sigma}_{2},\boldsymbol{\sigma}_{3}\}$ are the Pauli matrices. From Eqs.~(\ref{E})--(\ref{Stokes}) we obtain
\begin{equation}
\label{DoubleSlitStokes}
\begin{aligned}
\!\!\!S_j(\mathbf{r}) =& \; S_{j}^{\prime}(\mathbf{r})+S_{j}^{\prime\prime}(\mathbf{r})+2[ S_{0}^{\prime}(\mathbf{r})S_{0}^{\prime\prime}(\mathbf{r})]^{1/2} \\ & \times\!|s_j(\mathbf{r}_{1},\mathbf{r}_{2})|\cos(\alpha_{j}-k\Delta r), \ \ j\in\{0,1,2,3\},
\end{aligned}
\end{equation}
with $S_{j}^{\prime}(\mathbf{r})$ and $S_{j}^{\prime\prime}(\mathbf{r})$ being the Stokes parameters on $\mathcal{B}$ when only the slit at $\mathbf{r}_{1}$ or $\mathbf{r}_{2}$ is open, respectively, and $\Delta r=r_{1}-r_{2}$. Furthermore, $|s_{j}(\mathbf{r}_{1}, \mathbf{r}_{1})|$ and $\alpha_{j}$ are the magnitudes and phases of
\begin{equation}
\label{NormStokes}
s_{j}(\mathbf{r}_{1},\mathbf{r}_{2})=\frac{S_{j}(\mathbf{r}_{1},\mathbf{r}_{2})}{ [S_{0}(\mathbf{r}_{1})S_{0}(\mathbf{r}_{2})]^{1/2}}, \ \ j\in\{0,1,2,3\},
\end{equation} 
which in turn are the intensity-normalized versions of
\begin{equation}
\label{TwoPointStokes}
S_{j}(\mathbf{r}_{1},\mathbf{r}_{2})=\bra{\Psi}\!\hat{\mathbf{E}}^{\dagger}(\mathbf{r}_{1})\boldsymbol{\sigma}_{j}\hat{\mathbf{E}}(\mathbf{r}_{2})\!\ket{\Psi}, \ \ j\in\{0,1,2,3\}.
\end{equation} 
The quantities in Eq.~(\ref{TwoPointStokes}) are the coherence (two-point) Stokes parameters that contain all the information on the first-order vector-field correlations between $\mathbf{r}_{1}$ and $\mathbf{r}_{2}$ in the slit plane $\mathcal{A}$ \cite{Norrman2020,Korotkova2005,Tervo2009,Friberg2016}. The conventional (one-point) Stokes parameters at the slits are obtained from Eq.~(\ref{TwoPointStokes}) as $S_{j}(\mathbf{r}_{1},\mathbf{r}_{1})=S_{j}(\mathbf{r}_{1})$ and $S_{j}(\mathbf{r}_{2},\mathbf{r}_{2})=S_{j}(\mathbf{r}_{2})$.

\begin{figure}[b]
\includegraphics[width=.3  \textwidth]{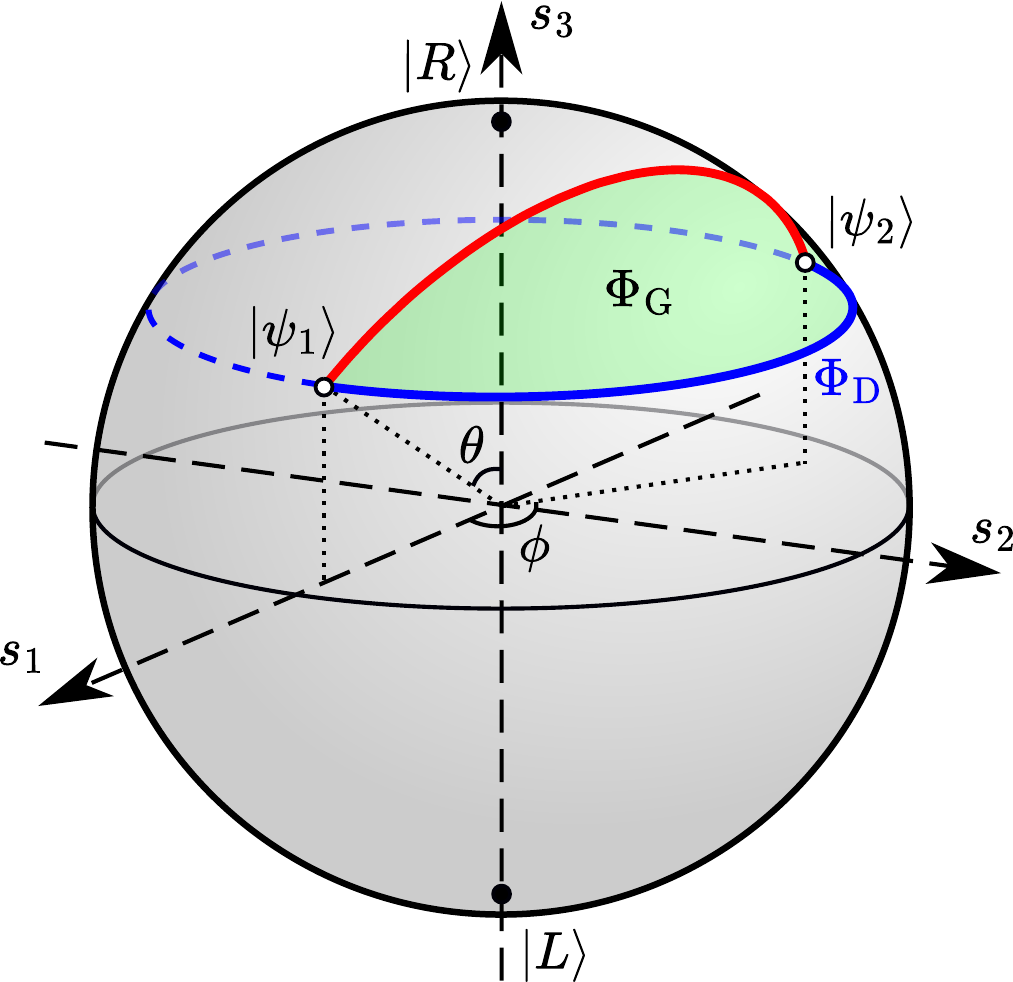}
\caption{Polarization-state evolution of the photon on the Poincar{\'e} sphere spanned by the intensity-normalized Stokes parameters $\{s_{1},s_{2},s_{3}\}$. The polarization states are specified by the spherical polar angles $0\leq\theta\leq\pi$ and $0\leq\phi\leq2\pi$, with $\theta=0$ and $\theta=\pi$ representing right-handed and left-handed polarization states $\ket{R}$ and $\ket{L}$, respectively. The blue curve represents the dynamical phase $\Phi_{\mathrm{D}}$ between the initial state $\ket{\psi_{1}}$ and final state $\ket{\psi_{2}}$, whereas the red curve is the geodesic between these states. The geometric phase $\Phi_{\mathrm{G}}$ equals half the green surface area enclosed by the curves.}
\label{fig2}
\end{figure}

Equation~(\ref{DoubleSlitStokes}) particularly shows that the polarization state (and intensity) on $\mathcal{B}$ varies periodically when moving the observation point $\mathbf{r}$ transversally along the screen. In addition, since any such one-photon light is completely first-order coherent at $\mathcal{A}$ \cite{Norrman2017} and thereby fully polarized on $\mathcal{B}$~\cite{Voipio2015}, so that $S_{1}^{2}(\mathbf{r})+S_{2}^{2}(\mathbf{r})+S_{3}^{2}(\mathbf{r})=S_{0}^{2}(\mathbf{r})$, the polarization-state evolution in the observation plane occurs on the surface of the associated Poincar\'e sphere (cf.\ Fig.~\ref{fig2})~\cite{Brosseau1998,Friberg2016}. These observations will be of central importance when we next assess the geometric phase on screen $\mathcal{B}$.

\emph{Geometric phase}.---To ascertain the geometric phase exhibited by the photon, we employ the quantum kinematic approach introduced by Mukunda and Simon \cite{Mukunda1993}. Let us consider a generic pure quantum state that evolves along some smooth curve $\gamma$ within the Hilbert space, i.e., $\ket{\psi(\gamma_1)}\rightarrow\ket{\psi(\gamma_2)}$, where $\ket{\psi(\gamma_1)}$ is the initial state and $\ket{\psi(\gamma_2)}$ is the final state. In this case, the geometric phase $\Phi_{\mathrm{G}}$ can be expressed as the difference
\begin{equation}
\label{Gphase} 
\Phi_{\mathrm{G}}=\Phi_{\mathrm{T}}-\Phi_{\mathrm{D}},
\end{equation}
where the first term
\begin{equation}
\label{Tphase} 
\Phi_{\mathrm{T}}=\arg\braket{\psi(\gamma_{1})|\psi(\gamma_{2})}
\end{equation} 
is the total phase between the initial and final states, and the second term
\begin{equation}
\label{Dphase} 
\Phi_{\mathrm{D}}=\mathrm{Im}\left[\int_{\gamma_1}^{\gamma_2}\bra{\psi(\gamma)}\!\frac{d}{d\gamma}\!\ket{\psi(\gamma)}d\gamma\right]
\end{equation}
is the dynamic phase along the path.

We recall from our above discussion that light in the pure single-photon state given by Eq.~(\ref{photon}) is fully polarized in the observation plane, with its polarization state evolving cyclically on the surface of the Poincar\'e sphere. It has further been shown that for any such polarized light field the evolution path has the form of a circle \cite{Hannonen2019}. We can therfore describe the polarization-state evolution of the photon on screen $\mathcal{B}$ in terms of the following qubit:
\begin{equation}
\label{Hstate}
\ket {\psi(\gamma)} = \cos(\theta/2)\ket{R}+e^{i\gamma\phi}\sin(\theta/2)\ket{L}.
\end{equation} 
Here $\theta\in[0,\pi]$ and $\phi\in[0,2\pi]$ are the polar and azimuthal angles on the Poincar\'e sphere as depicted in Fig.~\ref{fig2}, $\ket{R}$ and $\ket{L}$ are right-handed and left-handed circular polarization bases, respectively, and $\gamma\in[0,1]$ is a continuous real parameter. Our choice of a constant angle $\theta$ simplifies but does not reduce the generality of the analysis, since circular paths of any other orientation can always be transformed into the form of Eq.~(\ref{Hstate}) by a suitable rotation. The chosen value range for $\phi$, corresponding to one circular loop (cycle) on the Poincaré sphere, spans over a single spatial period in the interference pattern. We note, however, that the relation between $\phi$ and $\Delta r$ is in general not linear.

We can now connect the polarization-state evolution on the Poincar\'e sphere with the quantum kinematic approach. On substituting Eq.~(\ref{Hstate}) into Eqs.~(\ref{Gphase})--(\ref{Dphase}), with $\gamma_{1}=0$ and $\gamma_{2}=1$, we find that the geometric phase displayed by the photon in the observation plane $\mathcal{B}$ is
\begin{equation}  
\label{Gphasephoton1}
\begin{aligned}
\Phi_{\mathrm{G}}=& \arctan\left[\frac{\sin^{2}(\theta/2)\sin\phi}{\cos^{2}(\theta/2)+\sin^{2}(\theta/2)\cos\phi}\right] \\&-\frac{\phi}{2}\big(1-\cos\theta\big).
\end{aligned}
\end{equation}
Equation~(\ref{Gphasephoton1}) is fully general in the sense that it covers both cyclic ($\phi=2\pi$) and noncyclic ($\phi<2\pi$) evolution. It is formally similar to the classical geometric phase~\cite{Leinonen2023} and consistent with the so-called geodesic rule~\cite{Samuel1988,vanDijk2010a,vanDijk2010b,Zhou2020}: the start and end points in any noncyclic evolution should be connected with a geodesic, and the
geometric phase is specified by half the enclosed solid angle (surface area). Some important features can be concluded from Eq.~(\ref{Gphasephoton1}). The phase magnitude is bounded as $0\leq|\Phi_{\mathrm{G}}|\leq\pi$, and in any cyclic evolution it reduces to $|\Phi_{\mathrm{G}}|=\pi(1-\cos{\theta})$. For the special cases $\theta=0$ and $\theta=\pi$, corresponding to the north and south poles on the Poincar\'e sphere, respectively, we have $|\Phi_{\mathrm{G}}|=0$. Likewise, when the polarization path is on the equator, $\theta=\pi/2$, the geometric phase is zero for any trajectory within the range $0\leq\phi<\pi$, while it suddenly changes to $|\Phi_{\mathrm{G}}|=\pi$ for $\pi<\phi\leq 2\pi$.

\emph{Which-path information}.---To quantify the which-path information (WPI) carried by the photon at the two slits, we utilize the following two measures (applicable to any mixed state represented by a density operator $\hat{\rho}$)~\cite{Norrman2017}:
\begin{equation}  
\label{D}
D_{0}=\frac{|S_{0}(\mathbf{r}_{1})-S_{0}(\mathbf{r}_{2})|}{S_{0}(\mathbf{r}_{1})+S_{0}(\mathbf{r}_{2})}, \ \
D_{S}=\frac{|\mathbf{S}(\mathbf{r}_{1})-\mathbf{S}(\mathbf{r}_{2})|}{S_{0}(\mathbf{r}_{1})+S_{0}(\mathbf{r}_{2})}.
\end{equation}
The quantity $D_{0}$ is called the intensity distinguishability and it describes the intensity difference between the slits. The quantity $D_{S}$, with $\mathbf{S}(\mathbf{r}_{m})=[S_{1}(\mathbf{r}_{m}), S_{2}(\mathbf{r}_{m}), S_{3}(\mathbf{r}_{m})]$ in slit $m\in\{1, 2\}$, is the Stokes (or polarization) distinguishability that characterizes the polarization-state difference in the slit plane. The denominators in Eq.~(\ref{D}) ensure that $0\leq D_{0}\leq1$ and $0\leq D_{S}\leq1$.

For the single-photon state in Eq.~(\ref{photon}), the two general measures in Eq.~(\ref{D}) turn into
\begin{equation}
\label{WPI}
D_{0}=|p_{1}-p_{2}|, \ \ D_{S}=\sqrt{1-4|c_{1x}^{\ast}c_{2x}+c_{1y}^{\ast}c_{2y}|^{2}},
\end{equation}
where $p_{m}=|c_{mx}|^{2}+|c_{my}|^{2}$ is the path probability of the photon to pass via the slit $m\in\{1, 2\}$. At this one-photon level $D_{0}$ represents the path predictability~\cite{Englert1996,Norrman2017}, i.e., the possibility to correctly guess which of the slits the photon traverses based on its initial state preparation. For example, when $p_{1}\gg p_{2}$ the probability of detecting the photon in slit 1 is much larger than finding it in slit 2, yielding a high path predictability ($D_{0}\approx1$). In contrast, for $p_{1}\approx p_{2}$ the path predictability is negligible ($D_{0}\approx0$). Likewise, in this single-photon case $D_{S}$ translates into the path distinguishability~\cite{Englert1996,Norrman2017}, which describes one’s ability to discriminate the photon's path with respect to its polarization state at the two slits. Especially, maximum path distinguishability ($D_{S}=1$) is reached whenever the photon is orthogonally polarized in the slit plane. For instance, if we were to measure $x$-polarized ($y$-polarized) light when $c_{1y}=c_{2x}=0$, then a count signal in the detection plane would directly reveal that the photon has passed the first (second) slit. On the other hand, minimum path distinguishability ($D_{S}=0$) occurs if $\mathbf{S}(\mathbf{r}_{1})=\mathbf{S}(\mathbf{r}_{2})$. Generally, however, the WPI of the photon is characterized by partial path predictability ($0<D_{0}<1$) and distinguishability ($0<D_{S}<1$).

We further introduce the quantity
\begin{equation}
\label{WPIratio1}
d=\frac{D_{0}}{D_{S}},
\end{equation}
which is the ratio between the two different WPI species. For any polarized light, including the one-photon state of Eq.~(\ref{photon}), the two measures in Eq.~(\ref{D}) obey $D_{0}\leq D_{S}$~\cite{Norrman2017}. The WPI ratio in Eq.~(\ref{WPIratio1}) therefore satisfies $0\leq d\leq1$, physically meaning that the photon carries more WPI in terms of its polarization state than in terms of its path probabilities. The lower limit $d=0$ is reached only if the path predictability is zero ($D_{0}=0$), whereas the upper bound $d=1$ is saturated only for scalar light ($D_{0}=D_{S}$). In particular, for any polarized light field under cyclic evolution the geometric phase is related to the intensity and polarization-state differences at the slits according to $|\Phi_{\mathrm{G}}|=\pi[1-|S_{0}(\mathbf{r}_{1})-S_{0}(\mathbf{r}_{2})|/|\mathbf{S}(\mathbf{r}_{1})-\mathbf{S}(\mathbf{r}_{2})|]$~\cite{Hannonen2020,Hannonen2019}. On connecting this expression with Eqs.~(\ref{Gphasephoton1}) and (\ref{WPIratio1}), we then find that the polar angle $\theta$ in the Poincar{\'e} sphere representation is directly linked to the WPI ratio $d$ as
\begin{equation}
\label{WPIratio2}
d=\cos{\theta}.
\end{equation}
Equation~(\ref{WPIratio2}) covers also the noncyclic case as $\theta$ remains unaffected when varying the azimuthal angle $\phi$.

\emph{Wave--particle duality}.---We are now in a position to present the main result of this Letter. By using Eq.~(\ref{WPIratio2}) and standard trigonometric identities, we can first write the geometric phase in Eq.~(\ref{Gphasephoton1}) as
\begin{equation}
\label{Gphasephoton2}
\Phi_{\mathrm{G}}=\arctan\left(\frac{\sin{\phi}}{\eta+\cos{\phi}}\right)-\frac{\phi}{2}(1-d),
\end{equation} 
where $\eta=(1+d)/(1-d)$. We observe that the geometric phase in Eq.~(\ref{Gphasephoton2}) is now expressed solely in terms of the WPI ratio $d$ and the azimuthal angle $\phi$ (corresponding to the distance along the observation screen). Equation~(\ref{Gphasephoton2}) thus provides an exact link between the WPI ratio at the slits and the geometric phase in the detection plane.

By eventually introducing the $\pi$-normalized geometric phase $\Phi_{\mathrm{G}}^{\prime}=\Phi_{\mathrm{G}}/\pi$, which is bounded as $0\leq|\Phi_{\mathrm{G}}^{\prime}|\leq1$, we discover from Eq.~(\ref{Gphasephoton2}) the complementarity relation
\begin{equation}
\label{duality}
|\Phi_{\mathrm{G}}^{\prime}|+d\leq1.
\end{equation}
This result is obtained by first considering the maximum possible value of $|\Phi_{\mathrm{G}}^{\prime}|$ when $0<d<1$, and then separately analyzing the special scenarios of $d=0$ and $d=1$. When $0<d<1$, we find the maximum by taking the derivative of Eq.~(\ref{Gphasephoton2}) with respect to $\phi$. The corresponding zero yields $\phi=2\pi$, which stands for cyclic evolution. Substituting this value into Eq.~(\ref{Gphasephoton2}) results in a strict complementarity identity for any cyclic evolution, i.e.,
\begin{equation}
\label{dualitycyclic}
|\Phi_{\mathrm{G}}^{\prime}|+d=1, \ \ \mathrm{if} \ \phi=2\pi.
\end{equation}
For noncyclic evolution we then necessarily have
\begin{equation}
\label{dualitynoncyclic}
|\Phi_{\mathrm{G}}^{\prime}|+d<1, \ \ \mathrm{if} \ \phi<2\pi.
\end{equation}
The special case of $d=0$ is encountered solely if the path predictability is zero ($D_{0}=0$). In this situation Eq.~(\ref{Gphasephoton2}) leads to $\Phi_{\mathrm{G}}^{\prime}=-\lfloor(\phi+\pi)/2\pi\rfloor$, where $\lfloor\cdots\rfloor$ denotes the floor function. We thus find $|\Phi_{\mathrm{G}}^{\prime}|=0$ if $0\leq\phi<\pi$ and $|\Phi_{\mathrm{G}}^{\prime}|=1$ if $\pi<\phi\leq2\pi$, both of which satisfy Eq.~(\ref{duality}). The last case with $d=1$ is the trivial scalar-light scenario ($D_{0}=D_{S}$) for which Eq.~(\ref{Gphasephoton2}) directly gives $|\Phi_{\mathrm{G}}^{\prime}|=0$, which is also encompassed by Eq.~(\ref{duality}). 

Equation~(\ref{duality}) constitutes the main result of our work. It can be viewed as a fundamental quantificator of wave--particle duality of the photon in terms of the geometric
phase in the observation plane $\mathcal{B}$ (wave aspect) and the WPI ratio in the slit plane $\mathcal{A}$ (particle aspect). As underlined by Eq.~(\ref{dualitycyclic}), for cyclic evolution these two attributes are strictly mutually exclusive, i.e., reducing (increasing) $d$ increases (reduces) $|\Phi_{\mathrm{G}}|$ such that their combined sum equals exactly one.

\emph{Conclusions}.---In summary, we have explored the geometric phase of a single photon in the double-slit setup. In particular, we discovered a fundamental wave--particle duality relation for the photon that interconnects the geometric phase it exhibits in the observation plane and the which-path information it encompasses in the slit plane. This complementarity relation sets no restrictions on the polarization state at the slits and covers both cyclic and noncyclic polarization evolution in the interference plane. In the cyclic case, the general inequality turns into a tight identity which states that the geometric phase and which-path information are of strictly complementary nature. Due to their universal physical characters, we expect that similar features between the geometric phase and wave--particle duality exist in other quantum systems as well. Our work thereby unifies fundamenal notions in physics, reveals uncharted facets of the dual wave--particle nature of light, and identifies future directions towards research on the geometric phase.

\emph{Acknowledgments}.---The authors would like to thank Robert Fickler, Rafael Barros, and Jaime Moreno for fruitful discussions. This research was supported by the Research Council of Finland (Grant Nos.~354918, 349396, and 346518).

\end{document}